\documentclass[useAMS,usenatbib]{mn2e}
\usepackage{lmodern}

\usepackage[T1]{fontenc}
\usepackage[latin9]{inputenc}
\usepackage{color}
\usepackage{amsmath}
\usepackage{graphicx}
\usepackage{amssymb}
\usepackage[normalem]{ulem}
\usepackage{esint}
\usepackage[authoryear]{natbib}

\makeatletter
\pdfoutput=1

\usepackage[colorlinks,bookmarks]{hyperref}\definecolor{linkblue}{rgb}{0,0,0.8}
\definecolor{linkgreen}{rgb}{0,0.5,0}
\definecolor{darkgreen}{rgb}{0,0.4,0}
\definecolor{purple}{rgb}{0.7,0.0,0.4}
\hypersetup{linkcolor=linkblue, citecolor=purple, urlcolor=linkblue}

\providecommand{\eprint}[1]{\href{http://arxiv.org/abs/#1}{#1}}
\providecommand{\adsurl}[1]{\href{#1}{ADS}}

\usepackage{ifthen}\def\eprinttmp@#1arXiv:#2 [#3]#4@{
\ifthenelse{\equal{#3}{x}}{\href{http://arxiv.org/abs/#1}{#1}
}{\href{http://arxiv.org/abs/#2}{arXiv:#2} [#3]}}
\renewcommand{\eprint}[1]{\eprinttmp@#1arXiv: [x]@}

\newcommand{\Om}{\ensuremath{\Omega_\mathrm{m0}}\xspace}

\def\Om{$\Omega_m$}

\title[Breaking the spell of Gaussianity: higher order FM]{Breaking the spell of Gaussianity: forecasting with higher order Fisher matrices}
\author[E. Sellentin, M. Quartin and L. Amendola]{
Elena Sellentin$^{1}$,
Miguel Quartin$^{2}$,
Luca Amendola$^{1}$.\\
$^{1}$Institut F\"ur Theoretische Physik, Ruprecht-Karls-Universit\"at Heidelberg, Philosophenweg 16, 69120 Heidelberg, Germany\\
$^{2}$Instituto de Fisica, Universidade Federal do Rio de Janeiro, CEP,\\
21941-972, Rio de Janeiro, RJ, Brazil}

\makeatother

\begin{document}

\date{Accepted XXX. Received XXXX; in original form 02/04/2014.}

\maketitle
\pagerange{\pageref{firstpage}--\pageref{lastpage}} \pubyear{2014}

\label{firstpage}
\begin{abstract}
    We present the new method DALI (Derivative Approximation for LIkelihoods)
    for reconstructing and forecasting posteriors. DALI extends the Fisher
    Matrix formalism but allows for a much wider range of posterior shapes.
    While the Fisher Matrix formalism is limited to yield ellipsoidal
    confidence contours, our method can reproduce the often observed flexed,
    deformed or curved shapes of known posteriors. This gain in shape
    fidelity is obtained by expanding the posterior to higher order in
    derivatives with respect to parameters, such that non-Gaussianity
    in the parameter space is taken into account. The resulting expansion
    is positive definite and normalizable at every order. Here, we present
    the new technique, highlight its advantages and limitations, and show
    a representative application to a posterior of dark energy parameters
    from supernovae measurements.
\end{abstract}

\section{Introduction}

In the last few years the Fisher Matrix (FM) formalism has been widely
applied to forecast constraints on cosmological parameters from future
experiments~\citep[see e.g.][]{Tegmark:1996bz,Bassett:2009uv,wang10,2011JCAP...10..010B,
Abramo:2012,Debono:2013,Amendola:2013qna}.
With its recipe-like structure and its many elementary maths operations,
this technique knows how to entice scientists away from more complex
methods such as the Markov-Chain Monte Carlo~\citep[MCMC;][]{Christensen:2001gj,Lewis:2002ah,Dunkley:2004sv,Akeret:2012ky}),
nested sampling~\citep{Feroz:2007kg,Feroz:2008xx} or full-grid analysis~\citep{Tegmark:2000db},
although these other methods are known to reproduce the shape of posteriors
much more faithfully. The omnipresence of the FM is mainly caused
by its speedy execution. A fast posterior evaluation is indeed sometimes
more than a convenience: in~\citep{Amendola:2012wc,Heneka:2013hka}
a blind search for systematics on the Union2.1 supernova (SNeIa) data set~\citep{AmanullahLidman2010}
required roughly $10^{6}$ such evaluations, and the FM had to be
employed whenever valid.

While speed certainly is an important asset for a forecasting technique,
often one desires the essential shape of the posterior to be captured,
such that degeneracy directions and regions of the parameter space
that are not preferred by the data are represented adequately. In
this respect, the FM has often been criticized since it assumes that
the posterior is a Gaussian function of the parameters, and therefore
is bound to produce ellipsoidal confidence-level contours. Of these
ellipses, the principal axes represent the local direction of parameter
degeneracies, and the area of the ellipses is taken as a measure of
the constraining power of an experiment~\citep[Figure of Merit;][]{Albrecht:2006um,Euclid}.
However, a mismatch between both the orientation and the size of these
ellipses with respect to MCMC-generated posteriors has often been
observed~\citep{Wolz:2012,Khedekar:2013,Rodriguez:2013}. To
which extent Fisher Matrices are a trustworthy forecasting technique,
is consequently a debatable question.

These drawbacks of the FM originate from its assumption of the posterior being Gaussian in the parameters, which is exact only when the data are Gaussian and the model is linear in the parameters. This assumption is approximately true when one has collected enough data such that the central limit theorem kicks in.  However, it is often the case that the amount of data is insufficient to warrant such an approximation, except perhaps close to the maximum of the posterior. In fact, for many parameters of dark-energy-related research, targeted parameters such as $w_{a}$ (see definition below) are weakly constrained non-linear model parameters, such that the posterior contains a non-negligible amount of non-Gaussianity. Therefore, an obvious method to improve the description of the posterior beyond the scope of the FM is to tackle the Gaussian assumption. One recent investigation used invertible transformation of parameters in order to make the posterior more Gaussian~\citep{Joachimi:2011iq,2002PhRvD..66f3007K}.

Here instead we build on the FM, but expand the posterior to higher
orders. If the posterior $P$ really is Gaussian in the parameters,
the higher order derivatives of $\,\log P\,$ will be zero, such that
the extended method falls back onto the FM and nothing is lost. If
they are non-zero, a gain in shape fidelity is to be expected. As
many posteriors have a smooth shape and resemble often a {}``surrealistic''
version of an ellipse, i.e. the ellipses are slightly curved, flexed
or otherwise distorted, already the inclusion of just a few higher-order
derivatives promises good improvements.

The main problem in expanding over a Gaussian distribution is that
the expansion is in general not guaranteed to be a true distribution,
i.e. positive definite and normalizable. Edgeworth or Gram-Charlier
series suffer indeed from this serious problem. Here we find that
a simple rearrangement of the terms in the Taylor series can guarantee
that the expansion remains a true distribution at every order. The
expansion turns out to be a derivative expansion rather than a parameter
expansion, as we clarify below.

One of the most severe caveats of the FM is that it does not carry
any information that allows to check whether its assumption of (approximate)
Gaussianity is fulfilled. Therefore, one must rely on alternative
techniques if one is worried about the breakdown of the FM estimates.
Besides correcting the shapes of contours, higher order corrections
in the Taylor expansion of the posterior also serve as a fast (and
simplest) double-check on the Fisher Matrix analysis.

This paper is organized as follows: In an attempt to clearly separate
lengthy calculations from our argumentation line, we promote the appendices
to a vital part of the paper. In Sect.~\ref{sec:taylor} we develop
the extended formalism of posterior reconstruction and focus on a
parameter-independent covariance matrix. The derivative expansion
is discussed in Sect.~\ref{matrices}. In Sect.~\ref{sec:sn} we
specialize the method to SNeIa data and we apply it in Sect.~\ref{Dali-Union} to the Union2.1 supernova catalog and to a mock catalog with 1000 SNeIa up to $z=2$.
We then discuss in Sect.~\ref{marginalization} the issue of marginalization
of parameters. We conclude
in Sect.~\ref{sec:con}. The appendix contains in depth calculations,
a comparison between frequentist and Bayesian FM and an extension
of our method to parameter-dependent covariance matrices.

\section{Including Non-Gaussianity into a Posterior}\label{sec:Dali-method}

\subsection{Taylor Expansion of the Posterior}\label{sec:taylor}

We consider a posterior\footnote{Our results hold for a likelihood exactly in the way they hold for a posterior. This is especially true when we use uniform priors, since then the difference between likelihood and posterior in practice vanishes.} that depends on $n$ parameters $p_{\alpha}$,
where $\alpha$ can take values $1\,...\, n$. Denoting with $P$
the posterior distribution, we expand the log-likelihood $\mathcal{L}=-\log(\mathrm{P})$
as a function of the parameters $p_{\alpha}$ in Taylor series around
the likelihood best fit, indicated by the subscript~$0$ as
\begin{equation}
\begin{aligned}-\mathcal{L}\equiv\log\mathrm{P}\approx & \log\mathrm{P}_{0}+\frac{1}{2}\left.\big(\log\mathrm{P}\big)_{,\alpha\beta}\right|_{0}\;\Delta p_{\alpha}\Delta p_{\beta}\\
 & +\frac{1}{3!}\left.\big(\log\mathrm{P}\big)_{,\alpha\beta\gamma}\right|_{0}\;\Delta p_{\alpha}\Delta p_{\beta}\Delta p_{\gamma}\\
 & +\frac{{1}}{4!}\left.\big(\log\mathrm{P}\big)_{,\alpha\beta\gamma\delta}\right|_{0}\;\Delta p_{\alpha}\Delta p_{\beta}\Delta p_{\gamma}\Delta p_{\delta},\end{aligned}
\end{equation}
where summation over repeated indices is implied,       $\Delta p_{\alpha}=p_{\alpha}-\hat{p}_{\alpha}$
is the deviation of a parameter from its best-fitting $\,\hat{p}_{\alpha}$ and ${}_,\alpha \equiv \partial_{p_\alpha}$.
The first order derivatives vanish because we are at the maximum of
the posterior. Expanding to the second order yields the Fisher approximation.
From the third order onwards, non-Gaussianities are taken into account,
which correct for misestimates of the posterior by the FM, and thereby
lead to a deformation of its shape. We can write the approximation as
\begin{equation}
\begin{aligned}\mathrm{P}=N\exp\bigg[ & \left.-\frac{1}{2}F_{\alpha\beta}\Delta p_{\alpha}\Delta p_{\beta}-\frac{1}{3!}S_{\alpha\beta\gamma}\Delta p_{\alpha}\Delta p_{\beta}\Delta p_{\gamma}\right.\\
 & -\frac{1}{4!}Q_{\alpha\beta\gamma\delta}\Delta p_{\alpha}\Delta p_{\beta}\Delta p_{\gamma}\Delta p_{\delta}-\,{\cal O}(5)\,\bigg]\,,\end{aligned}
\label{eq:exp}\end{equation}
 where $N$ is a normalization constant, \begin{equation}
\begin{aligned}F_{\alpha\beta} & =\mathcal{L}_{,\alpha\beta}\,,\\
S_{\alpha\beta\gamma} & =\mathcal{L}_{,\alpha\beta\gamma}\,,\\
Q_{\alpha\beta\gamma\delta} & =\mathcal{L}_{,\alpha\beta\gamma\delta}\,,\end{aligned}
\label{eq:def-ffq}\end{equation}
 and we shall neglect the fifth and higher order terms of the Taylor
series. From now on all derivatives in the Taylor series are taken
at the best-fitting value.

Here, the $n\times n$ matrix $F_{\alpha\beta}$ is the usual FM. We dub the $n\times n\times n$ tensor $S_{\alpha\beta\gamma}$
the Flexion\footnote{We borrow the term {}``flexion'' from the weak-lensing literature~\citep{Goldberg:2004hh,Bacon:2005qr},
where it also refers to third order corrections to the shapes of images,
which typically flex the shape of sources from ellipses towards a
banana-shaped image.%
} tensor and the scalar \begin{equation}
S\equiv S_{\alpha\beta\gamma}\Delta p_{\alpha}\Delta p_{\beta}\Delta p_{\gamma}\label{eq:flexion-S}\end{equation}
 just {}`the Flexion'. Likewise we call $Q_{\alpha\beta\gamma\delta}$
the Quarxion tensor and \begin{equation}
Q\equiv Q_{\alpha\beta\gamma\delta}\Delta p_{\alpha}\Delta p_{\beta}\Delta p_{\gamma}\Delta p_{\delta}\label{eq:quarxion-Q}\end{equation}
 just {}`the Quarxion'. Finally, for simplicity we dub \begin{equation}
F\equiv F_{\alpha\beta}\Delta p_{\alpha}\Delta p_{\beta}\label{eq:fisher-F}\end{equation}
 just {}`the Fisher'. We therefore refer to the expansion Eq.~\eqref{eq:exp}
up to fourth order as to the Fisher-Flexion-Quarxion approximation. Any non-zero Flexion or Quarxion tensor implies immediately
that the posterior is not exactly Gaussian in the parameters, and
the larger their components are, the larger is the non-Gaussianity.

In the frequentist approach, the FM is defined as the data
average of $F_{\alpha\beta}$, i.e. \begin{equation}
F_{\alpha\beta}^{F}\equiv\langle\mathcal{L}{}_{,\alpha\beta}\rangle\,.\end{equation}
 In the Bayesian approach the data are no longer random variables
and no averaging takes place. We have instead the alternative definition
\begin{equation}
F_{\alpha\beta}^{B}\equiv\left.\mathcal{L}{}_{,\alpha\beta}\right|_{BF}\,,\end{equation}
 that is, the F Mis evaluated at the parameter maximum-likelihood
best fit. This point is sometimes neglected in the literature and
in Appendix~\ref{app:freqbay} we comment on the difference between
these two definitions. Nevertheless, when making a forecast for a
future experiment, the maximum likelihood parameter set is chosen
beforehand (it is the fiducial set) and the two definitions coincide.
In this paper we assume the frequentist definition because it allows
for several simplifications and because the whole Fisher approach
(and the extension here proposed) is most useful when doing forecasts.

Note that the exponential in the FM approximation contains only a
quadratic form. The argument of the exponential function is consequently
always negative, which ensures that the probability stays finite.
This handy feature is not necessarily true for the Quarxions and never
true for the Flexions: The Flexion is cubic in the $\Delta p$ and
will therefore always become negative at large enough $\Delta p$.
Whenever negative Flexion and Quarxions terms become larger than the
Fisher, the argument of the exponential becomes positive and the Fisher-Flexion-Quarxion approximation
diverges at large $\Delta p$. This is a fundamental problem in many
expansions around a Gaussian, such as the Edgeworth or the Gram-Charlier.

It is however possible to solve this problem by expanding in derivatives
rather than in $\Delta p$, as we show next.

\subsection{DALI: The Derivative Expansion}\label{matrices}

We consider now cases in which the parameters appear only in a theoretical
model $\mu$ that is compared to a data set, and not in the covariance
matrix of the parameter space. We label the theoretical prediction
corresponding to the $i$-th data point as $\mu_{i}$; notation can
be simplified by introducing the model vector $\boldsymbol{\mu}$.
In this paper, Latin indices generally run over the data and Greek
indices over the parameters.

Averaging over possible data sets generated from a given fiducial,
we find that the Fisher Matrix is given by (see Appendix~\ref{app:ffq}),
\begin{equation}
F_{\alpha\beta}=\langle\mathcal{L}_{,\alpha}\mathcal{L}_{,\beta}\rangle\label{eq:freqfm}\end{equation}
 i.e.~no second derivatives appear. With $M=C^{-1}$ being the inverse of the parameter-independent and positive-definite
covariance matrix in the data space, we find in Appendix~\ref{app:ffq}
the Flexion tensor to be \begin{equation}
\begin{aligned}S_{\alpha\beta\gamma} & =\langle\mathcal{L}{}_{,\alpha\beta}\mathcal{L}_{,\gamma}\rangle+\langle\mathcal{L}{}_{,\gamma\alpha}\mathcal{L}_{,\beta}\rangle+\langle\mathcal{L}{}_{,\beta\gamma}\mathcal{L}_{,\alpha}\rangle\\
 & =\boldsymbol{\mu}_{,\alpha\beta}M\boldsymbol{\mu}_{,\gamma}+\mathrm{cycl}\,.,\end{aligned}
\label{eq:flexion}\end{equation}
 The Quarxion tensor is \begin{equation}
\begin{aligned}Q_{\alpha\beta\gamma\delta} & =\boldsymbol{\mu}_{,\alpha\gamma\delta}M\boldsymbol{\mu}_{,\beta}+\boldsymbol{\mu}_{,\delta\gamma}M\boldsymbol{\mu}_{,\beta\alpha}\\
 & +\boldsymbol{\mu}_{,\alpha\beta\delta}M\boldsymbol{\mu}_{,\gamma}+\boldsymbol{\mu}_{,\delta\beta}M\boldsymbol{\mu}_{,\gamma\alpha}\\
 & +\boldsymbol{\mu}_{,\alpha\gamma\beta}M\boldsymbol{\mu}_{,\delta}+\boldsymbol{\mu}_{,\beta\gamma}M\boldsymbol{\mu}_{,\delta\alpha}\\
 & +\boldsymbol{\mu}_{,\delta\gamma\beta}M\boldsymbol{\mu}_{,\alpha}\,.
 \end{aligned}
\label{eq:quarxion}\end{equation}
 It is obvious from their definition in equation (\ref{eq:def-ffq}) that Flexions
and Quarxions are symmetric under index permutation. Both Flexion
and Quarxion tensors also transform under parameter-space transformations
in the same way as the FM: to wit, with a series of simple Jacobian
transformations. When taking the full Flexion or Quarxion term, all the distinct terms
of the same type in Eqs. \eqref{eq:flexion} and \eqref{eq:quarxion} become
indistinguishable. For instance, by renaming the indices,
\begin{align}
\boldsymbol{\mu}_{,\alpha\beta}M\boldsymbol{\mu}_{,\gamma}\Delta p_{\alpha}\Delta p_{\beta}\Delta p_{\gamma} & =\boldsymbol{\mu}_{,\gamma\alpha}M\boldsymbol{\mu}_{,\beta}\Delta p_{\gamma}\Delta p_{\alpha}\Delta p_{\beta} \\
= & \boldsymbol{\mu}_{,\beta\gamma}M\boldsymbol{\mu}_{,\alpha} \Delta p_{\beta}\Delta p_{\gamma}\Delta p_{\alpha} \,.
\end{align}
Therefore we can simplify
\begin{equation}
\begin{aligned}
S & =3\boldsymbol{\mu}_{,\alpha\beta}M\boldsymbol{\mu}_{,\gamma}\Delta p_{\alpha}\Delta p_{\beta}\Delta p_{\gamma}\,,\\
Q & =(4\boldsymbol{\mu}_{,\alpha\gamma\delta}M\boldsymbol{\mu}_{,\beta}+3\boldsymbol{\mu}_{,\delta\gamma}M\boldsymbol{\mu}_{,\beta\alpha})\Delta p_{\alpha}\Delta p_{\beta}\Delta p_{\gamma}\Delta p_{\delta}\,.
\end{aligned}
\label{eq:simplified}
\end{equation}

Although some terms in $Q$ are positive definite (e.g. $\boldsymbol{\mu}_{,\delta\gamma}M\boldsymbol{\mu}_{,\beta\alpha}$),
it appears that neither $S$ nor $Q$ are globally positive definite;
this problems shows up at all orders. However, as anticipated, the
expansion can be arranged also in a different way, namely in order
of derivatives. That is, to second order in the $\boldsymbol{\mu}$
derivatives we have
\begin{equation}
\begin{aligned}\mathrm{P}=N\exp\bigg[ & -\frac{1}{2}\boldsymbol{\mu}_{,\alpha}M\boldsymbol{\mu}_{,\beta}\Delta p_{\alpha}\Delta p_{\beta} \\
& - \bigg(\frac{1}{2}\boldsymbol{\mu}_{,\alpha\beta}M\boldsymbol{\mu}_{,\gamma}\Delta p_{\alpha}\Delta p_{\beta}\Delta p_{\gamma}\\
 & +\frac{1}{8}\boldsymbol{\mu}_{,\delta\gamma}M\boldsymbol{\mu}_{,\beta\alpha}\Delta p_{\alpha}\Delta p_{\beta}\Delta p_{\gamma}\Delta p_{\delta}\bigg) +\,{\cal O}(3)\,\bigg].
 \end{aligned}
\label{eq:exp-1}\end{equation}
The decisive advantage of this expression is that now the expansion
is a true distribution, i.e. normalizable and positive definite, since
the highest-order term  in $\Delta p$, $\boldsymbol{\mu}_{,\delta\gamma}M\boldsymbol{\mu}_{,\beta\alpha}\Delta p_{\alpha}\Delta p_{\beta}\Delta p_{\gamma}\Delta p_{\delta}=(\boldsymbol{\mu}_{,\delta\gamma}\Delta p_{\gamma}\Delta p_{\delta})^{2}M$,
is positive-definite (if, as we assumed from the start, the data inverse
covariance matrix $M$ is itself positive definite). Remarkably,
this is true at every order; for instance, at third order we have
\begin{equation}
\begin{aligned}
    \mathrm{P} & = N \\
    &\exp \bigg[ \! -\frac{1}{2}\boldsymbol{\mu}_{,\alpha}M\boldsymbol{\mu}_{,\beta}\Delta p_{\alpha}\Delta p_{\beta}
     -\bigg(\!\frac{1}{2}\boldsymbol{\mu}_{,\alpha\beta}M\boldsymbol{\mu}_{,\gamma}\Delta p_{\alpha}\Delta p_{\beta}\Delta p_{\gamma}\\
    & + \frac{1}{8}\boldsymbol{\mu}_{,\delta\gamma}M\boldsymbol{\mu}_{,\beta\alpha}\Delta p_{\alpha}\Delta p_{\beta}\Delta p_{\gamma}\Delta p_{\delta}\bigg)\\
    & -\bigg(\frac{1}{6}\boldsymbol{\mu}_{,\delta}M\boldsymbol{\mu}_{,\beta\alpha\gamma}\Delta p_{\alpha}\Delta p_{\beta}\Delta p_{\gamma}\Delta p_{\delta}\\
    &+\frac{1}{3!2!}\boldsymbol{\mu}_{,\alpha\beta\delta}M\boldsymbol{\mu}_{,\gamma\tau}\Delta p_{\alpha}\Delta p_{\beta}\Delta p_{\gamma}\Delta p_{\delta}\Delta p_{\tau}\,\\
    & +\frac{1}{3!3!2!}\boldsymbol{\mu}_{,\alpha\beta\gamma}M\boldsymbol{\mu}_{,\delta\tau\sigma}\Delta p_{\alpha}\Delta p_{\beta}\Delta p_{\gamma}\Delta p_{\delta}\Delta p_{\tau}\Delta p_{\sigma}\!\!\bigg)\! + {\cal O}(4)\!\bigg]\!,
 \end{aligned}
\label{eq:exp-1-1}\end{equation}
 where again one sees that the leading term, the last one in Eq.~\eqref{eq:exp-1-1},
is positive definite.
Notice that the derivative expansion requires only derivatives of order $N/2$ (for $N$ even) or $(N+1)/2$ (for $N$ odd) for an expansion of order $N$ in $\Delta p$, rather than $N-1$ as in the expansion (\ref{eq:exp}). The numerical coefficient for a term of order $N$ in $\Delta p$ formed with $n_{1}$ and $\,n_{2}=N-n_{1}\,$ derivatives is $\,(n_{1}!n_{2}!)^{-1}\,$ for $\,n_{1}\not=n_{2}\,$ and $\,[2(n_{1}!)^{2}]^{-1}$ for $n_{1}=n_{2}\,$ (see Appendix~\ref{app:dependent}).

The approximated posteriors \eqref{eq:exp-1}--\eqref{eq:exp-1-1} are the main product of this paper: they represent true distributions and the second- and third-derivative correction, respectively, over
the Fisher approximation. We baptize this new posterior reconstruction method DALI: Derivative Approximation for LIkelihoods. For the sake of clear referencing, we further call the approximation Eq.~(\ref{eq:exp-1}) in which the leading terms are second derivatives the ``doublet-DALI'' and the
approximation that has third derivatives as leading order [Eq.~(\ref{eq:exp-1-1})] the {}``triplet-DALI''.

The derivative expansion can actually be directly obtained in a very
simple way. We label the $i$-th data point of the data set by $m_{i}$,
and combine them into a vector $\mathbf{m}$. Let us start from the
standard Gaussian likelihood exponent
\begin{equation}
\frac{1}{2}\big[\boldsymbol{m}-\boldsymbol{\mu}(p_{1},...,p_{n})\big] M\big[\boldsymbol{m}-\boldsymbol{\mu}(p_{1},...,p_{n})\big].\label{eq:orig-1}
\end{equation}
Now we expand to second order around the best fit $\hat{p}_{\alpha}$
\begin{equation}
\boldsymbol{\mu}\approx\hat{\boldsymbol{\mu}}+\boldsymbol{\mu}_{,\alpha}\Delta p_{\alpha}+\frac{1}{2}\boldsymbol{\mu}_{,\alpha\beta}\Delta p_{\alpha}\Delta p_{\beta}\,,
\end{equation}
 where $\hat{\mu}\equiv\mu(\hat{p}_{1},...,\hat{p}_{n})$, so we obtain
\begin{equation}
\begin{aligned}
\frac{1}{2}&\big[\boldsymbol{m}  -\boldsymbol{\mu}(p_{1},...,p_{n})\big]  M\big[\boldsymbol{m}-\boldsymbol{\mu}(p_{1},...,p_{n})\big]\approx\\
 & \frac{1}{2}[\boldsymbol{m}-\hat{\boldsymbol{\mu}}]M[\boldsymbol{m}-\hat{\boldsymbol{\mu}}]\\
 & -\big(\boldsymbol{m}-\hat{\boldsymbol{\mu}}\big)M\Big(\boldsymbol{\mu}_{,\alpha}\Delta p_{\alpha}+\frac{1}{2}\boldsymbol{\mu}_{,\alpha\beta}\Delta p_{\alpha}\Delta p_{\beta}\Big)\\
 & +\frac{1}{2}\boldsymbol{\mu}_{,\alpha}M\boldsymbol{\mu}_{,\beta}\Delta p_{\alpha}\Delta p_{\beta}+\frac{1}{2}\boldsymbol{\mu}_{,\alpha}M\boldsymbol{\mu}_{,\beta\gamma}\Delta p_{\alpha}\Delta p_{\beta}\Delta p_{\gamma}\\
 & +\frac{1}{8}\boldsymbol{\mu}_{,\alpha\beta}M\boldsymbol{\mu}_{,\gamma\delta}\Delta p_{\alpha}\Delta p_{\beta}\Delta p_{\gamma}\Delta p_{\delta}\,.
\end{aligned}
\end{equation}
 The first term on the rhs is an irrelevant constant that can be absorbed
in the normalization; the second term averages out to zero, while
the remaining terms are indeed as in Eq.~(\ref{eq:exp-1}). It is
worth remarking again that the expansions Eq.~(\ref{eq:exp}) and
Eq.~(\ref{eq:exp-1}) are mathematically equivalent; it is only when
arranged in order of derivatives rather than in powers of $\Delta p$ that
they differ at each finite order.

In Appendix~\ref{app:dependent} we extend this formalism to parameter-dependent
correlations. We leave however tests of this case in realistic cosmological
scenarios to future work.

\begin{figure*}
    \centering \includegraphics[width=1\textwidth]{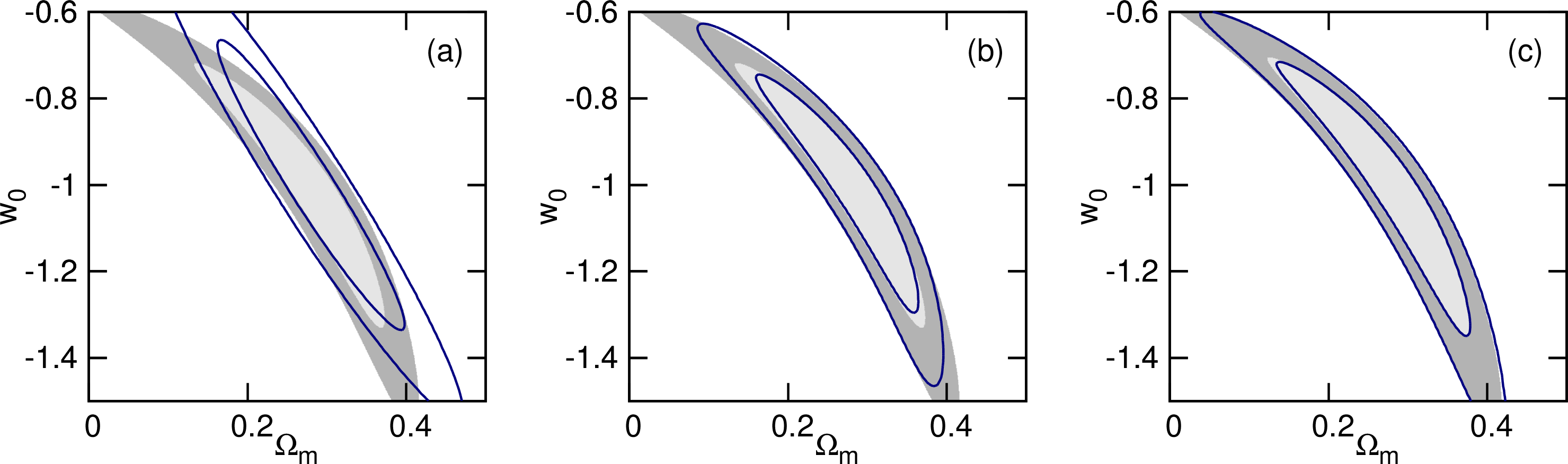} \caption{Comparison of the full, non-approximated posterior of the SNeIa Union2.1 catalogue (grey) with different approximations (dark-blue). In this plot only we fix $w_a = 0$ (i.e., assume what is often called ``$w$CDM'' model).     The confidence contours are drawn at the 1 and 2$\sigma$ confidence levels. Panel (a): The Fisher Matrix approximation; panel (b): Eq.~(\ref{eq:exp-1}), the doublet-DALI approximation of the posterior includes well the non-Gaussianities; panel~(c): Eq.~(\ref{eq:exp-1-1}), the triplet-DALI approximation captures the non-Gaussianities even better. \label{EvolutionPan1}}
\end{figure*}

\subsection{Speed and complexity}

The one incontrovertible advantage of the FM is its speed. A quick
order of estimate of the complexity of the
DALI approximation can be obtained by observing that the expensive computations needed for the matrices are the evaluations of the vectors of the derivatives. For $n$ parameters, there are $n$ possible first derivatives, so the complexity rises linear with $n$. For the `doublet' correction Eq.~(\ref{eq:exp-1}) one needs also the second derivatives, of which there are $n(n+1)/2$ distinct ones for $n$ parameters,
and similarly $(n^3 + 3n^2+2n)/6$ for the `triplet' correction.  Since every numerical derivative of order $p$ requires (at lowest accuracy) $p+1$ evaluations of the posterior, the  complexity for large $n$ goes like
$n^2$ and $(2/3) n^3$ for the doublet and triplet, respectively. In comparison, grids or MCMC routines evaluate the full likelihood (which implies generating theoretical predictions of the data at every point in parameter space) typically thousands of times already for e.g. four parameters. Therefore only for ${\cal O}(1000)$ [${\cal O}(100)$] parameters does the doublet [triplet] require roughly the same ${\cal O}(10^{6})$ evaluations of a typical Monte Carlo run in large parameter spaces.
In practice the evaluation of the posterior is thus significantly faster with DALI, as most forecasts in cosmology rely on less than dozen free parameters, and the posterior can be numerically costly to compute. Note however, that only Gaussian posteriors are again Gaussians with less dimensions if they are marginalized. This analytical result makes marginalizations with FM extremely fast. For non-Gaussian posteriors, for which DALI is interesting, there exists no general analytical marginalization. Therefore DALI will be slower in this respect than FM - a price that one has to pay, if the non-Gaussianity of a posterior shall be captured.

\section{DALI Method at Work}\label{work}

\subsection{Specialization to Supernovae}\label{sec:sn}

We consider now an application of our method to SNeIa data. The
measurable quantity is the distance modulus, which is related to the
dimensionless luminosity distance by, \begin{equation}
\mu_{i}=5\log\hat{d}(z_{i}),\end{equation}
 where the index $i$ denotes the dependence on a given redshift.
The likelihood function for the supernovae after marginalization of
the Hubble constant and the absolute luminosity is~\citep{Amendola:2010} \begin{equation}
\mathcal{L}=-\log L=\frac{1}{2}\left(S_{2}-\frac{S_{1}^{2}}{S_{0}}\right),\end{equation}
 where the sums are \begin{equation}
S_{n}=\sum_{i}\frac{(m_{i}-\mu_{i})^{n}}{\sigma_{i}^{2}},\label{eq:sum}\end{equation}
 where $m_{i}$ is a measurement at redshift $z_{i}$ and the corresponding
theoretical mean $\mu_{i}$. The log-likelihood can be written as
\begin{equation}
\mathcal{L}=\frac{1}{2}X_{i}M_{ij}X_{j},\end{equation}
 where $X_{i}=m_{i}-\mu_{i}$ and the inverse covariance matrix is
\begin{equation}
M_{ij}=s_{i}s_{j}\delta_{ij}-\frac{s_{i}^{2}s_{j}^{2}}{S_{0}},\label{eq:Mij-full}\end{equation}
 (no sum) where $s_{i}=1/\sigma_{i}$. If one assumes $s_{i}=1/\sigma$
(constant) then the covariance matrix is \begin{equation}
M_{ij}=\sigma^{-2}\left(\delta_{ij}-\frac{1}{N}\right).\label{eq:M}\end{equation}
 So finally we have \begin{equation}
\begin{aligned}F_{\alpha\beta}^{\text{SN}} & =\left\langle \left(\frac{\partial\mu_{i}}{\partial p_{\alpha}}M_{ij}X_{j}\right)^{2}\right\rangle ,\\
 & =25\,\frac{\partial\log\hat{d}_{i}}{\partial p_{\alpha}}M_{ij}\frac{\partial\log\hat{d}_{j}}{\partial p_{\beta}}\,.\end{aligned}
\end{equation}
 Similarly, the Flexion and Quarxion tensors and the DALI expansion
are then obtained by replacing $\mu_{i}$ with $5\log\hat{d}_{i}$.

Note that a parameter that appears additively in $\mu_i$, like the offset, will not enter the DALI terms; therefore, the analytic marginalization of the posterior affects only the Fisher term and remains analytic also in DALI.

\subsection{Applying DALI to the supernova catalogues}\label{Dali-Union}

In order to demonstrate the potential of DALI, we show how accurately
it can recover the {}``banana-shaped'' posterior of the supernova
Union2.1 catalogue~\citep{AmanullahLidman2010}. This catalogue comprises
the distance moduli of 580 SNeIa, which we use for the data points
$m_{i}(z_i)$ of Eq.~(\ref{eq:sum}), together with their respective errors
$\sigma_{i}$. We compare this data set with the distance moduli obtained
from a flat $w$CDM cosmology with the Chevallier-Polarski-Linder parametrization
for the dark energy equation of state~\citep{Chevallier:2000qy,Linder_wa} as
\begin{equation}
w(a)=w_{0}+w_{a}(1-a).
\end{equation}
We choose the fiducial parameters to be the best fit parameters of the SNeIa
posterior found in~\citep{AmanullahLidman2010} for the $w$CDM model and evaluate the distance moduli at the redshifts of the Union2.1 catalogue.

In Figure~\ref{EvolutionPan1} we depict in grey solid contours the non-approximated posterior (obtained with a grid method), which we will frequently refer to as the ``full'' posterior. Here and in all other figures the contours are drawn at 1 and 2$\sigma$ (we follow standard procedure and use $1\sigma$ and $2\sigma$ as shorthand notation for $68.3$ and $95.4\%$ confidence levels).
The improvement of the shape fidelity by successively adding higher order derivatives to the posterior can be seen as one inspects panel (a) [FM], (b) [doublet-DALI] and (c) [triplet-DALI] of Figure~\ref{EvolutionPan1}.

As the observational campaigns for dark energy proceed and more data are collected the posteriors
are expected to become more and more Gaussian. To investigate the use of DALI
in this respect, we mock a future supernova catalogue with 1000 SNeIa,
uniformly distributed in the redshift range $0<z<2$. We use a flat CPL-cosmology with $\Omega_m^{\rm fid}=0.285$, $w_{0}^{\rm fid}=-1$
and $w_{a}^{\rm fid}=0$ as fiducial. In the $\{$\Om$,\, w_{0}\}$--plane,
such a catalogue yields a posterior of similar shape to Figure~\ref{EvolutionPan1},
which the doublet- and triplet-DALI can recover nicely.

\begin{figure*}
\includegraphics[width=1\textwidth]{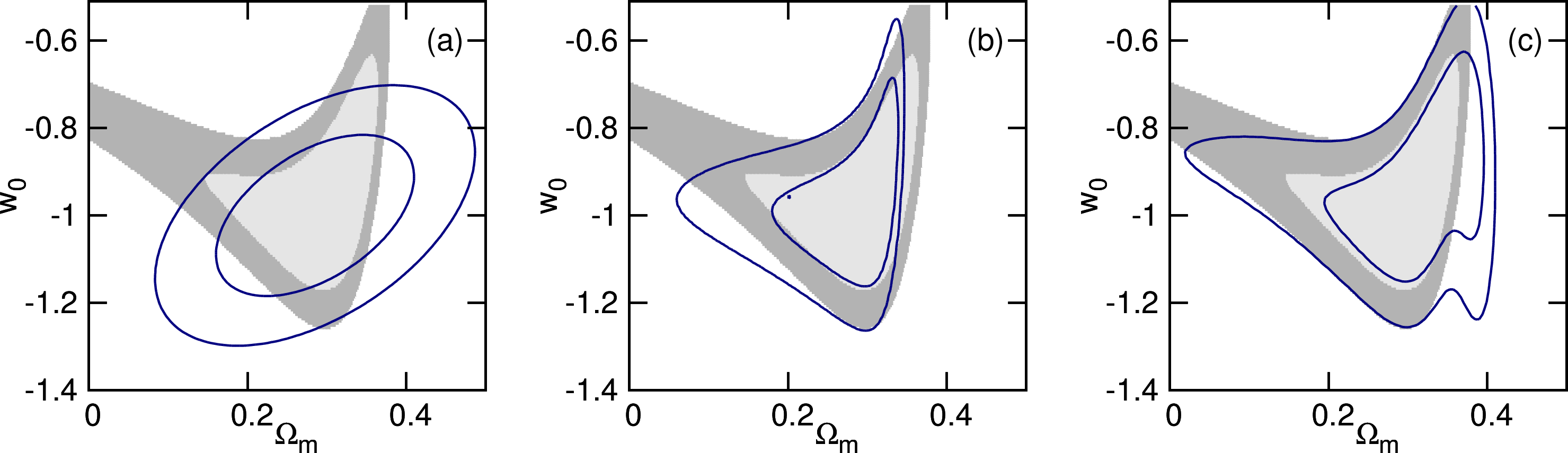}
\caption{Same as Figure~\ref{EvolutionPan1} but for the mock catalogue of 1000 SNeIa (see text) and marginalizing over $w_{a}$, which results in a heavily non-Gaussian grey posterior. Again the DALI methods capture the shape of the posterior much better than the Fisher Matrix. Note that the doublet-DALI is a very good compromise between speed and shape accuracy.
\label{nike}}
\end{figure*}
\begin{figure*}
\includegraphics[width=1\textwidth]{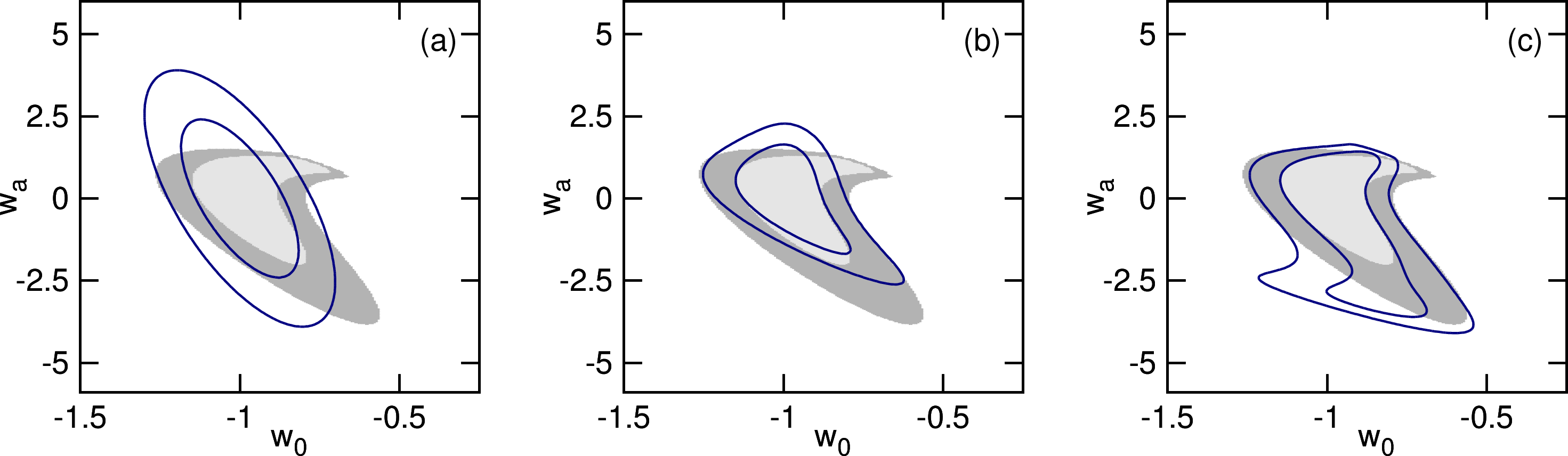}
\caption{Same as Figure~\ref{nike} but this time marginalized over $\Omega_{m}$
in the interval $[0,1]$.
Note that the upper half of the Fisher-ellipse covers parameter ranges with high $w_{a}$.  This indicates that Fisher does not capture the underlying physics well. In both derivative expansions, the posterior does not cover these regions.
\label{Badquarx}}
\end{figure*}
\begin{figure*}
\includegraphics[width=1\textwidth]{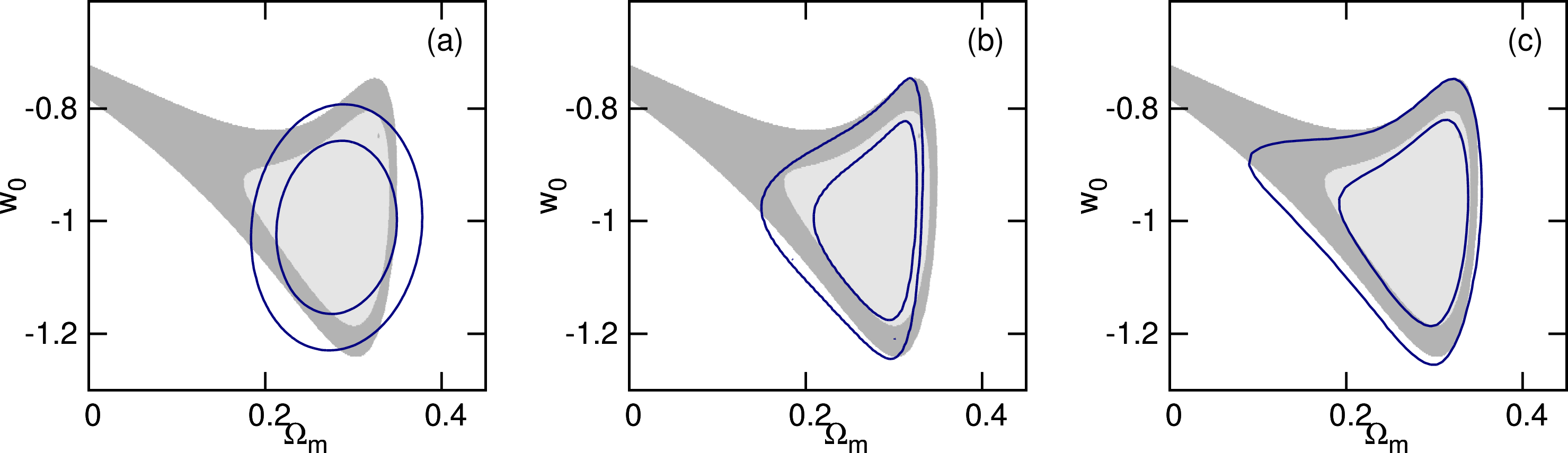}
\caption{Same as Figure~\ref{nike} but marginalized with a Gaussian prior of $\sigma_{w_a} = 1.0$, instead of a flat prior. For this smoother prior, the triplet-Dali contours do not leak out of the grey underlying posterior shape.
\label{Cens2Gauss}}
\end{figure*}
\begin{figure*}
\includegraphics[width=1\textwidth]{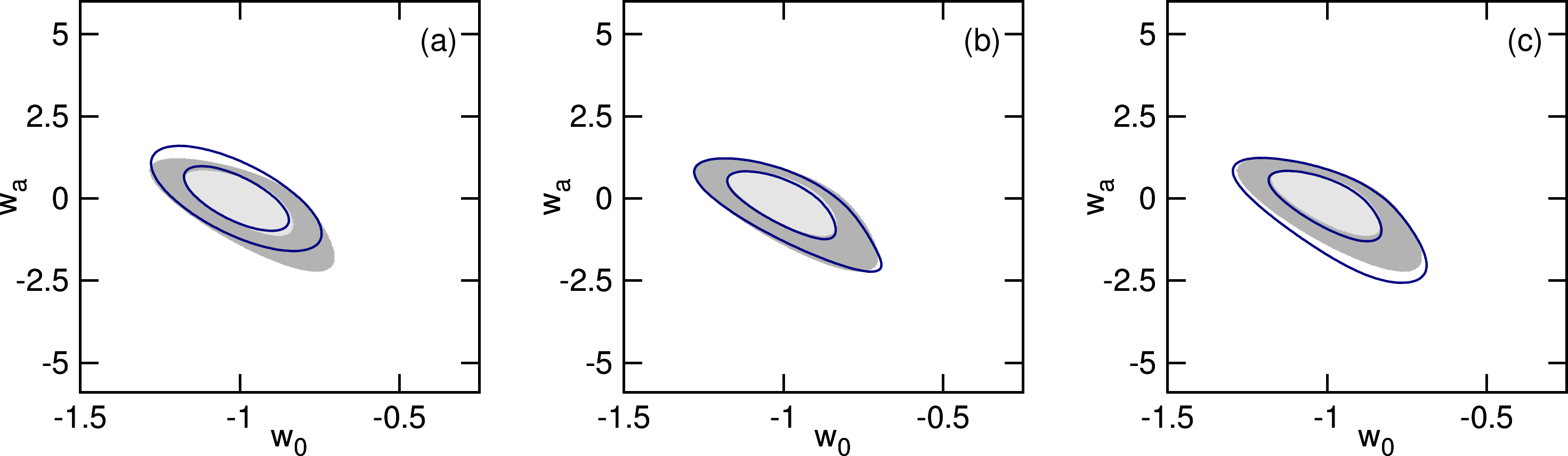}
\caption{Same as Figure~\ref{Badquarx} but marginalized with a Gaussian prior of $\sigma_{\Omega_m} = 0.03$, instead of a flat prior. Also for this rather tight prior, the marginalization leads to a noticeable amount of non-Gaussianity, which can be well captured by DALI.
\label{Cens3Gauss}}
\end{figure*}

We further demonstrate the potential of our method on a posterior with higher non-Gaussianity: in Figure~\ref{nike}, we marginalized the posterior of Figure~\ref{EvolutionPan1} over $w_{a}$ in the range $(-\infty,+\infty)$, and our method can recover the shape of this heavily non-Gaussian posterior quite accurately. In Figure~\ref{Badquarx}, we marginalized instead over \Om. Note that the upper half of the Fisher ellipse extends far into the the parameter space of positive $w_a$, which corresponds to a completely different expansion history of the universe, one that is ruled out at many $\sigma$ by supernova data. Changing from flat priors to Gaussian priors also affects the shapes of posteriors. We therefore show in Figure~\ref{Cens2Gauss} and in Figure~\ref{Cens3Gauss} that the DALI contours improve, as expected, with the marginalization over two particular cases of Gaussian priors: the former Gaussian in $w_a$, the latter in $\Omega_m$ (in each case we keep the priors on the remaining variables uniform).

\subsection{Marginalization}\label{marginalization}

The Fisher Matrix has four very useful properties: (i) it allows
one to evaluate the $n$-$\sigma$ confidence-level contours (which
in that case are just ellipses) analytically; (ii) the Gaussian approximation allows one to trivially achieve marginalization over parameters by dropping lines and columns
from its inverse matrix; (iii) fixing parameters at their best fit values
is similarly achieved by dropping the corresponding column and line
from the Fisher Matrix; (iv) the FM  of the product of two posteriors is the sum of the two posteriors' FMs. Only the last two properties are   shared
by the DALI method. The other two must be dealt with numerically.

In the vast majority of cases, we are interested in one or two-dimensional
contour plots of the posterior marginalized in all other parameters.
Marginalizations in the DALI must be carried out numerically
in an $n-$dimensional space. This is clearly a disadvantage
of the DALI method when compared to standard Fisher Matrix, as without any further simplifications the numerical
complexity will grow with the number $n$ of parameters in the
same way as in standard numerical integrations, which can be based
on either grids or Monte Carlo methods. Nevertheless since the needed
derivatives are only evaluated at the best fit (the fiducial), marginalizations can be carried out without evaluating the posterior, i.e. without running over the data for each parameter set. As discussed above in Sect.~\ref{matrices}, this makes the DALI method much faster than standard grids or (except for a \emph{very large} number of parameters) MCMC's.

\section{Conclusions}\label{sec:con}

Our new DALI method of posterior reconstruction was
developed to eliminate the drawback of the Fisher Matrix approach,
while making only small concessions in manners of speed. We achieved
this goal by expanding the posterior up to second or third order in
parameter derivatives, such that the approximation comprises a significant
amount of the non-Gaussianity in the parameter space. The new terms
give a fast measure of how much non-Gaussianity the posterior contains
and how accurately the Fisher Matrix reproduces the posterior. The
gain in shape fidelity when using the DALI method results in a more
faithful reconstruction of the posterior.

As an additional application, the DALI method could help MCMC routines to determine beforehand the
high-probability regions to explore. The speed of MCMC methods in fact have been known to be dependent on the shape of the so-called proposal distribution from which the random walk steps are selected. Usually, a simple multivariate Gaussian distribution is used, based on the FM expansion~\citep{Dunkley:2004sv}. Another option is to run a first crude MCMC-run and use the rough posterior estimate as a proposal distribution~\citep{Lewis:2002ah}. The DALI method offers a third alternative, one which we expect to allow for faster convergence than a simple multi-variate Gaussian.

\begin{figure}
\includegraphics[width=0.47\textwidth]{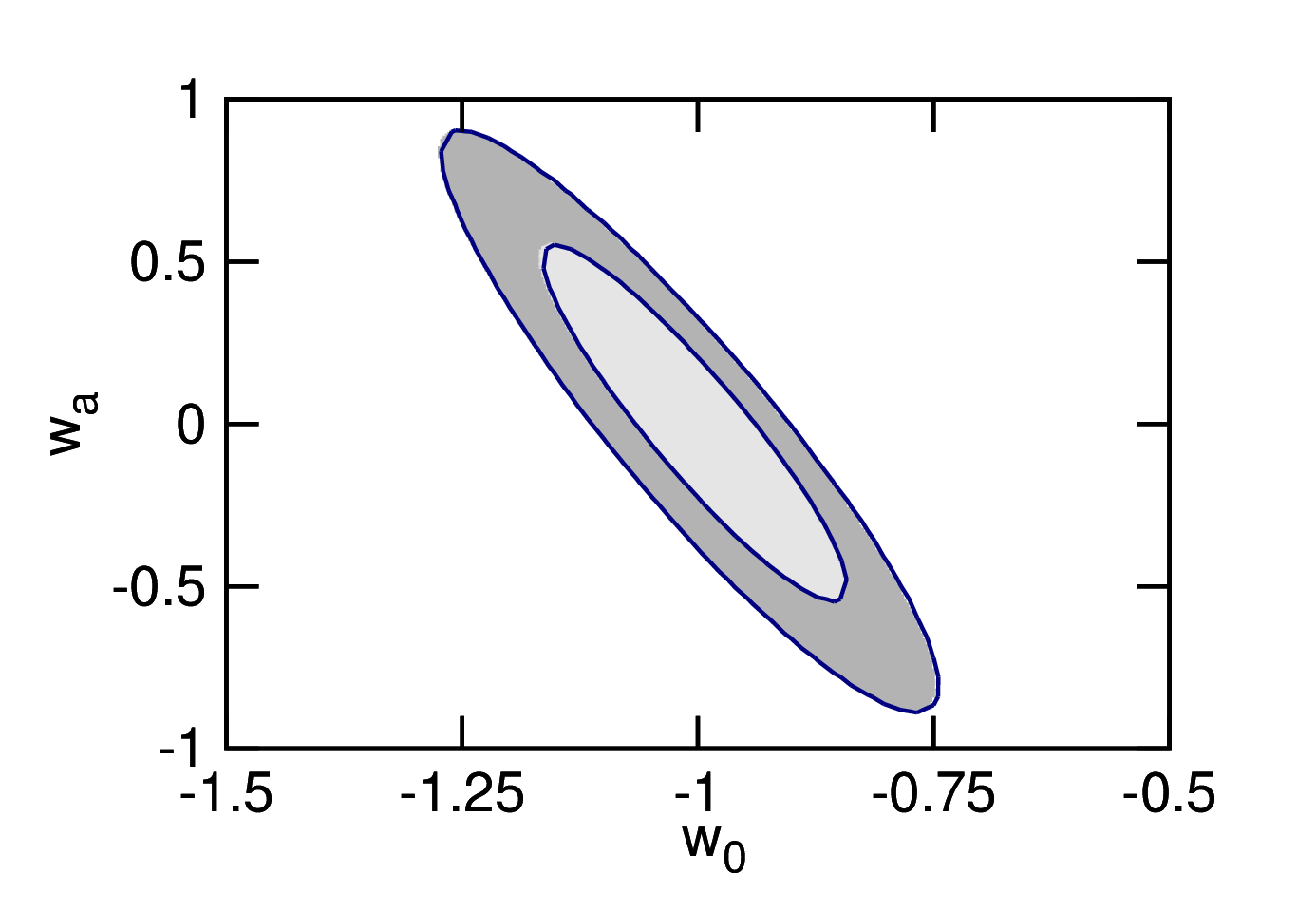}
\caption{Same as Figures~\ref{nike} and \ref{Badquarx} but for a fixed value of $\Omega_m = \Omega_m^{\rm fid}$.
If the posterior is highly Gaussian, the DALI method falls back onto
the FM and ellipsoidal confidence contours emerge.
\label{fig3}}%
\end{figure}

The DALI method can also be employed to gauge quantitatively how good is the FM approximation of Gaussianity. For example the posterior is very ellipsoidal in the parameters $w_{a}$ and $w_{0}$, as can be seen in Figure~\ref{fig3}. The DALI method then falls back on to the FM - with the important advantage of having checked that the assumption of a Gaussian posterior is justified. In fact, the FM in itself contains no information that allows one to carry out such a check, and authors sometimes run a full MCMC in order to compare the final contours~\citep{Wolz:2012,Rodriguez:2013}. A full MCMC run is obviously a somewhat costly numerical procedure (and may involve some trial-and-error), which although completely justifiable for final forecasts in expensive surveys is often not the most convenient one when fast results are desired. We nevertheless leave a more detailed exploration of how to best use the DALI method as a measurement of non-Gaussianity for future work.

Needless to say, the range of applicability of the DALI method is
not restricted to cosmology and can be applied to any data set.

\section*{Acknowledgement}

We give special thanks to Shuo Yuan for pointing out a mistake in our original Appendix A.
We would also like to thank Michael Hobson, Marcos Lima, Claudia Quercellini
and Jochen Weller for useful discussions. We thank Salvador Dali for
inspiration. LA and ES acknowledge financial support from DFG through
the TRR33 project {}``The Dark Universe''. MQ is grateful to Brazilian
research agencies CNPq and FAPERJ for support.

\appendix

\section{Frequentist and Bayesian Fisher matrix }

\label{app:freqbay}

Suppose there exists an observable $\mathbf{m}=[m_{1},...,m_{n}]$
to which a theoretical prediction by a model $\mu$ corresponds that
is a function of a parameter set: $\boldsymbol{\mu}=\boldsymbol{\mu}(p_{1},...,p_{m})$.
In the FM formalism the observed outcome is the mean values
of the observables assumed as the null hypothesis. This method allows
a quick way to estimate errors on cosmological parameters, given errors
in observable quantities. The FM is defined as the Hessian
of the log-likelihood function $\mathcal{L}=-\log(\mathrm{P})$, \begin{equation}
F_{\alpha\beta}=\left<-(\log{\mathrm{P}(\mathbf{m},\boldsymbol{\mu})),_{\alpha\beta}}\right>.\label{eq:freqfisher}\end{equation}
 This can be simplified as follows \begin{equation}
\begin{aligned}F_{\alpha\beta} & =\left<-(\log{\mathrm{P}(\mathbf{m},\boldsymbol{\mu})),_{\alpha\beta}}\right>\\
 & =\left<-\frac{\mathrm{P}_{,\alpha\beta}}{\mathrm{P}}+(\log\mathrm{P})_{,\alpha}(\log\mathrm{P})_{,\beta}\right>\\
 & =\left<(\log\mathrm{P})_{,\alpha}(\log\mathrm{P})_{,\beta}\right>\end{aligned}
\label{eq:freqfisher2}\end{equation}
 since \begin{equation}
\left<\frac{\mathrm{P}_{,\alpha\beta}}{\mathrm{P}}\right>
=\int\frac{\mathrm{P}_{,\alpha\beta}}{\mathrm{P}}\mathrm{P}d^{n}x
=\partial_{\alpha\beta}\int\mathrm{P}d^{n}x=0\,.
\end{equation}
 In the case of Gaussian data, the likelihood for $n$ data is \begin{equation}
\mathrm{P}=\frac{1}{(2\pi)^{n/2}\sqrt{|C|}}e^{-\frac{1}{2}(m_{i}-\mu_{i})C_{ij}^{-1}(m_{j}-\mu_{j})}\,,\end{equation}
 (notice that in this appendix we use the covariance matrix $C$
rather than its inverse $M$ as in the main text). The FM is then~\citep[suppressing the data indices $i,j$ and implicitly summing over them
unless otherwise specified;][]{Tegmark:1996bz} \begin{equation}
F_{\alpha\beta}=\frac{1}{2} \mathrm{Tr}\big[C_{,\alpha}C^{-1}C_{,\beta}C^{-1}\big]+
\boldsymbol{\mu}_{,\alpha}C^{-1}\boldsymbol{\mu}_{,\beta}\,.\end{equation}

By the Cramer-Rao inequality, a model parameter $p_{\alpha}$ cannot
have a variance smaller than $1/(F_{\alpha\alpha})^{1/2}$ (evaluated
for unbiased estimators) when all other parameters are fixed, or a
precision $(F^{-1})_{\alpha\alpha}^{1/2}$ when all other parameters
are marginalized over. Note however that the Cramer-Rao inequality
concerns variances and does not say anything about the relative size
of the confidence regions.

The purely Bayesian definition of the FM is instead:
\begin{equation}
F_{\alpha\beta}^{B}=
-\log(\mathrm{P}(\mathbf{m},\boldsymbol{\mu})),_{\alpha\beta}\big|_{BF}
\,,\label{eq:bayesfisher}\end{equation}
 where the derivatives have to be evaluated at the best fit values
of the parameters, i.e. for parameters such that \begin{equation}
\mathrm{P},_{\alpha}=0\,.\label{eq:bf}\end{equation}
 This definition makes no reference to the average over the data,
which in the Bayesian context are fixed once and for all by the current
experiment. Expressions \eqref{eq:freqfisher} and \eqref{eq:bayesfisher}
are however in general different and the Cramer-Rao inequality does
not hold in general for $F^{B}$. We can also write \begin{equation}
\begin{aligned}F_{\alpha\beta}^{B} & =\left.-\log(\mathrm{P}(\mathbf{m},\boldsymbol{\mu})),_{\alpha\beta}\right|_{BF}\\
 & =\left.-\frac{\mathrm{P}_{,\alpha\beta}}{\mathrm{P}}\right|_{BF}+\left.(\log\mathrm{P})_{,\alpha}(\log\mathrm{P})_{,\beta}\right|_{BF}\\
 & =\left.-\frac{\mathrm{P}_{,\alpha\beta}}{\mathrm{P}}\right|_{BF}\end{aligned}
\label{eq:bayesfisher-2}\end{equation}
 due to Eq.~\eqref{eq:bf} .

We show now that the only cases in which \eqref{eq:freqfisher} (evaluated on the best fit parameters) and \eqref{eq:bayesfisher} coincide are (\emph{a}) when the data are Gaussian and the parameters enter in a linear way in the mean and in the variance and (\emph{b}) in the case of forecasting.

In fact we have
\begin{equation}
    (\log\mathrm{P})_{,\alpha}=-\frac{1}{2}\mathrm{Tr}\big[C_{,\alpha}C^{-1}+ C^{-1}D_{,\alpha}-C^{-1}C_{,\alpha}C^{-1}D\big]\,,
\end{equation}
where we defined the  matrix
\begin{equation}
    D_{ij}=X_{i}X_{j}
\end{equation}
and the  vector
\begin{equation}
    \mathbf{X}\equiv\mathbf{m}-\boldsymbol{\mu}\,.
\end{equation}
The best fit condition $(\log L)_{,\alpha}=0$ gives
\begin{equation}
    \mathrm{Tr}[C^{-1}C_{,\alpha}C^{-1}D]=\mathrm{Tr}[C_{,\alpha}C^{-1}+C^{-1}D_{,\alpha}]\,.\label{eq:bfc}
\end{equation}
If $C$ does not depend on the parameters, the best fit equation
becomes
\begin{equation}
    D_{,\alpha}=0\,.
\end{equation}

In order to proceed we now draw attention to the fact that
\begin{align}
    {\rm Tr}\big[D_{,\alpha} \big] = -2 \mu_{i,\alpha} X_i\,.
\end{align}
The above allow us to write
\begin{equation}
\begin{aligned}
    F_{\alpha\beta}^{B} & \,=\,-(\log P)_{,\alpha\beta}|_{BF}\\
    & \,=\, \frac{1}{2} \mathrm{Tr}\big[C^{-1}C_{,\alpha}C^{-1}C_{,\beta}\big]+\boldsymbol{\mu}_{,\alpha}C^{-1}\boldsymbol{\mu}_{,\beta}+\Sigma_{\alpha\beta}\,,
\end{aligned}
\end{equation}
where
\begin{equation}
\begin{aligned}
    \Sigma_{\alpha\beta} \,\equiv\,  \frac{1}{2} \,
    \mathrm{Tr}\Big[& C^{-1}C_{,\alpha\beta}(I-C^{-1}D) -2 \,\mathbf{X} C^{-1} \boldsymbol{\mu}_{,\alpha\beta} \\
    & -2C_{,\alpha}C^{-1}C_{,\beta}C^{-1} - C^{-1}C_{,\beta}(C^{-1}D)_{,\alpha} \\
    &- C^{-1}C_{,\alpha}(C^{-1}D)_{,\beta}\Big].
\end{aligned}
\end{equation}

The matrix $\Sigma$ expresses the difference between frequentist and Bayesian FM. The first one is the one that ensures the Cramer-Rao inequality. The second one is the matrix that approximates the posterior.

Now, when we do forecasts, we generate mock data with variance given by $C$ and mean given by $\boldsymbol{\mu}$. If we evaluate the average FM for many mock data and note that
\begin{align}
    \langle D\rangle & = C \,,\\
    \langle \mathbf{X}\rangle & = 0 \,, \\
    \langle D_{ij}\rangle_{,\alpha} & =-\mu_{i,\alpha}\langle X_{j}\rangle-\langle X_{i}\rangle\mu_{j,\alpha}=0 \,,\\
    \langle D_{ij}\rangle_{,\alpha\beta} & = \langle-\mu_{i,\alpha\beta}X_j+ 2\mu_{i,\alpha}\mu_{j,\beta}- X_i\mu_{j,\alpha\beta}\rangle = 2 \mu_{i,\alpha}\mu_{j,\beta} ,
\end{align}
we obtain
\begin{equation}
    \langle\Sigma_{\alpha\beta}\rangle=0\,.
\end{equation}
Then in doing a forecast we in general identify the two FMs, or rather we can say that the generation of mock data implements the frequentist approach. Analysing real data, however, one should use the Bayesian FM, because this is the approximation to the posterior.

\section{Parameter Independent Covariance Matrix}

\label{app:ffq}

We assume in this Appendix that the parameters appear only in the
theoretical model $\mu$ that is compared to a data set. The data
covariance matrix shall be independent of parameters. In Appendix~\ref{app:dependent}
we extend our formalism to parameter-dependent correlations. Latin
indices run over the data, Greek index over the parameters.

Averaging over possible data sets generated from a given fiducial (subscript
0) we have \begin{equation}
\begin{aligned}F_{\alpha\beta} & \equiv\langle\mathcal{L}{}_{,\alpha\beta}\rangle_{0}\,,\\
S_{\alpha\beta\gamma} & \equiv\langle\mathcal{L}{}_{,\alpha\beta\gamma}\rangle{}_{0}\,,\\
Q_{\alpha\beta\gamma\delta} & \equiv\langle\mathcal{L}{}_{,\alpha\beta\gamma\delta}\rangle{}_{0}\,.\end{aligned}
\end{equation}
 Using the identities \begin{equation}
\begin{aligned}\langle\mathcal{L}_{,\alpha}\rangle & =0\,,\\
\Big\langle\frac{\mathrm{P}_{,\alpha\beta}}{\mathrm{P}}\Big\rangle & =0\,,\\
\Big\langle\frac{\mathrm{P}_{,\alpha\beta\gamma}}{\mathrm{P}}\Big\rangle & =0\,,\end{aligned}
\end{equation}
 we can show (see eq.~\ref{eq:freqfisher2}) that \begin{equation}
F_{\alpha\beta}=\langle\mathcal{L}_{,\alpha}\mathcal{L}_{,\beta}\rangle\,,\end{equation}
 so that no second derivatives appear. Note that \begin{equation}
\frac{\mathrm{P}_{,\alpha\beta}}{\mathrm{P}}=-\mathcal{L}{}_{,\alpha\beta}+\mathcal{L}_{,\alpha}\mathcal{L}_{,\beta}\,.\end{equation}
 The Flexion tensor is then \begin{equation}
\begin{aligned}S_{\alpha\beta\gamma} & =-\Big\langle\frac{\mathrm{P}_{,\alpha\beta\gamma}}{\mathrm{P}}\Big\rangle+\left(\Big\langle\frac{\mathrm{P}_{,\alpha\beta}\mathrm{P}_{,\gamma}}{P^{2}}\Big\rangle+\mathrm{cycl}\right)-2\Big\langle\frac{\mathrm{P}_{,\alpha}\mathrm{P}_{,\beta}\mathrm{P}_{,\gamma}}{\mathrm{P}^{3}}\Big\rangle\\
 & =\left(\langle\mathcal{L}{}_{,\alpha\beta}\mathcal{L}_{,\gamma}-\mathcal{L}_{,\alpha}\mathcal{L}_{,\beta}\mathcal{L}_{,\gamma}\rangle+\mathrm{cycl}\right)+2\langle\mathcal{L}_{,\alpha}\mathcal{L}_{,\beta}\mathcal{L}_{,\gamma}\rangle\\
 & =\left(\langle\mathcal{L}{}_{,\alpha\beta}\mathcal{L}_{,\gamma}\rangle+\mathrm{cycl}\right)-\langle\mathcal{L}_{,\alpha}\mathcal{L}_{,\beta}\mathcal{L}_{,\gamma}\rangle\,.\end{aligned}
\end{equation}
 We can make further progress by invoking the functional shape of
the log-likelihood \begin{equation}
\mathcal{L}={\rm const}+\frac{1}{2}(-\log{\normalcolor \textcolor{blue}{{\normalcolor \det}}M}+X_{i}M_{ij}X_{j})\,,\label{llike}\end{equation}
 where $M=C^{-1}$ is the inverse of the covariance matrix in the
parameter space.

If the parameters are only in $\boldsymbol{\mu}_{}$ we have \begin{align}
\mathcal{L}_{,\alpha} & =-\boldsymbol{\mu}_{,\alpha}M\mathbf{X}_{}\,,\label{eq:lamu}\\
\mathcal{L}_{,\alpha\beta} & =-\boldsymbol{\mu}_{,\alpha\beta}M\mathbf{X}_{}+\boldsymbol{\mu}_{,\alpha}M\boldsymbol{\mu}_{,\beta}\,.\label{eq:labmu}\end{align}
 When taking the data averages (denoted by $\langle\rangle$), all
odd powers of $X_{i}$ give zero and, since the data are Gaussian,
\begin{align}
\langle X_{j}X_{m}\rangle & =M_{jm}^{-1}\,,\\
\langle X_{i}X_{j}X_{\ell}X_{m}\rangle & =M_{ij}^{-1}M_{\ell m}^{-1}+M_{i\ell}^{-1}M_{jm}^{-1}+M_{im}^{-1}M_{\ell j}^{-1}\,,\\
\langle X_{i}X_{j}X_{\ell}X_{m}X_{k}X_{n}\rangle & =M_{ij}^{-1}M_{\ell m}^{-1}M_{kn}^{-1}+\mathrm{dist.\; perm.}\nonumber \\
&=15~{\rm {terms}\,,}
\end{align}
 (where only the distinguishable permutations have to be counted,
i.e. permutations that produce identical terms, e.g. $M_{ij}$ and
$M_{ji}$, must be discarded). This means that \begin{equation}
\langle\mathcal{L}_{,\alpha}\mathcal{L}_{,\beta}\mathcal{L}_{,\gamma}\rangle=0\,,\end{equation}
 and the Flexions matrix follows to be \begin{equation}
\begin{aligned}S_{\alpha\beta\gamma} & =\langle\mathcal{L}{}_{,\alpha\beta}\mathcal{L}_{,\gamma}\rangle+\mathrm{cycl}\\
 & =\mu_{i,\alpha\beta}M_{ij}\mu_{k,\gamma}M_{km}\langle X_{j}X_{m}\rangle+\mathrm{cycl}\\
 & =\boldsymbol{\mu}_{,\alpha\beta}M\boldsymbol{\mu}_{,\gamma}+\mathrm{cycl}\,.\end{aligned}
\end{equation}
 The Quarxions can be easily calculated from Eq.~(\ref{llike}) and
turn out to be, \begin{align}
\begin{split}Q_{\alpha\beta\gamma\delta} & =\big\langle\mathcal{L},_{\alpha\beta\gamma\delta}\big\rangle\\
 & =\boldsymbol{\mu}_{,\alpha\gamma\delta}M\boldsymbol{\mu}_{,\beta}+\boldsymbol{\mu}_{,\delta\gamma}M\boldsymbol{\mu}_{,\beta\alpha}\\
 & +\boldsymbol{\mu}_{,\alpha\beta\delta}M\boldsymbol{\mu}_{,\gamma}+\boldsymbol{\mu}_{,\delta\beta}M\boldsymbol{\mu}_{,\gamma\alpha}\\
 & +\boldsymbol{\mu}_{,\alpha\gamma\beta}M\boldsymbol{\mu}_{,\delta}+\boldsymbol{\mu}_{,\beta\gamma}M\boldsymbol{\mu}_{,\delta\alpha}\\
 & +\boldsymbol{\mu}_{,\delta\gamma\beta}M \boldsymbol{\mu}_{,\alpha}- \big\langle\boldsymbol{\mu}_{\alpha\beta\gamma\delta}M\boldsymbol{X}\big\rangle\,.
 \end{split}
\label{Alle}\end{align}
 The last term averages out to zero due to the Gaussian data, such
that no fourth order derivatives survive and what we are left with
for the Quarxions is Eq.~(\ref{eq:quarxion}).

\section{Parameter dependent covariance matrix}

\label{app:dependent}

If the parameters enter also the data covariance matrix $M$, we
have, instead of Eq. (\ref{eq:lamu}), \begin{equation}
\begin{aligned}\mathcal{L}_{,\alpha} & =-\frac{1}{2}T_{\alpha}-\mu_{i,\alpha}M_{ij}X_{j}\\
 & +\frac{1}{2}X_{i}M_{ij,\alpha}X_{j}\,,\end{aligned}
\end{equation}
 where we define \[
T_{\alpha}\equiv\mathrm{Tr}(M^{-1}M_{,\alpha})\]
 and instead of Eq. (\ref{eq:labmu}) for the second derivatives \begin{equation}
\begin{aligned}\mathcal{L}_{,\alpha\beta} & =-\frac{1}{2}\mathrm{Tr}(M^{-1}M_{,\alpha\beta}-M^{-1}M_{,\alpha}M^{-1}M_{,\beta})\\
 & -\mu_{i,\alpha\beta}M_{ij}X_{j}+\mu_{i,\alpha}M_{ij}\mu_{j,\beta}\\
 & +\frac{1}{2}X_{i}M_{ij,\alpha\beta}X_{j}-(\mu_{i,\alpha}M_{ij,\beta}+\mu_{i,\beta}M_{ij,\alpha})X_{j}\,.\end{aligned}
\end{equation}
 With a further derivative we obtain, in explicit notation, \begin{equation}
\begin{aligned}S_{\alpha\beta\gamma} & =\Big[\mu_{i,\alpha\beta}M_{ij}\mu_{j,\gamma}+\mu_{i,\alpha}M_{ij,\beta}\mu_{j,\gamma}\\
 & +\frac{1}{2}M_{ij,\alpha\beta}M_{\ell m,\gamma}M_{i\ell}^{-1}M_{jm}^{-1}+\mathrm{cycl.}\Big]\\
 & -M_{ij,\alpha}M_{k\ell,\beta}M_{mn,\gamma}M_{ik}^{-1}M_{jm}^{-1}M_{\ell n}^{-1}\,.\end{aligned}
\label{eq:c3}\end{equation}
 If $\mu_{i}=0$, for instance when applying the formalism to density
contrasts, then the Flexion tensor reduces to \begin{equation}
\begin{aligned}S_{\alpha\beta\gamma} & =\Big[\frac{1}{2}M_{ij,\alpha\beta}M_{\ell m,\gamma}M_{i\ell}^{-1}M_{jm}^{-1}+\mathrm{cycl.}\Big]\\
 & -M_{ij,\alpha}M_{k\ell,\beta}M_{mn,\gamma}M_{ik}^{-1}M_{jm}^{-1}M_{\ell n}^{-1}\,.\end{aligned}
\label{eq:c4}\end{equation}

For the Quarxions, the result in tensor notation is\begin{equation}
\begin{aligned}& Q_{\alpha\beta\gamma\delta}  =\\
& \;\; \Big[\mathbf{\boldsymbol{\mu}}_{,\alpha\beta\gamma}M\boldsymbol{\mu}_{,\delta}+\boldsymbol{\mu}_{,\alpha\beta}M\boldsymbol{\mu}_{,\gamma\delta}+\boldsymbol{\mu}_{,\alpha\beta}M_{,\gamma}\boldsymbol{\mu}_{,\delta}+\boldsymbol{\mu}_{,\alpha}M_{,\beta\gamma}\boldsymbol{\mu}_{,\delta}\\
 &\; \; +\frac{1}{2}M_{,\alpha\beta\gamma}M^{-1}M_{,\delta}M^{-1}-M_{,\alpha\beta}M^{-1}M_{,\gamma}M^{-1}M_{,\delta}M^{-1}\\
 &\;\; +\frac{1}{2}M_{,\alpha\beta}M^{-1}M_{,\gamma\delta}M^{-1}+\mathrm{dist.\; perm.}\Big]\\
 &\;\; +3M_{,\alpha}M^{-1}M_{,\beta}M^{-1}M_{,\gamma}M^{-1}M_{,\delta}M^{-1}\,.\end{aligned}
\end{equation}
 Here, $\mathrm{dist.\; perm.}$ means all the distinguishable permutations
(in Eqs.~\eqref{eq:c3}, \eqref{eq:c4} they coincide with cyclic permutations).
For instance, among all the possible permutations of the term $\mathbf{\boldsymbol{\mu}}_{,\alpha\beta\gamma}M\boldsymbol{\mu}_{,\delta}$
those that exchange $\alpha\beta\gamma$ give back the same term and
are to be neglected: in this case, the possible $4!=24$ permutations
of $\alpha\beta\gamma\delta$ reduce by a factor of $3!=6$ (the permutations
of $\alpha\beta\gamma$), leaving only 4 terms, as in Eq. (\ref{Alle}).
Similarly, for the term $\boldsymbol{\mu}_{,\alpha}M_{,\beta\gamma}\boldsymbol{\mu}_{,\delta}$,
the two permutations of $\alpha\delta$ and the two of $\beta\gamma$
are to be neglected, leaving 24/4=6 distinct permutations. For the
terms without derivatives of $M$ and $n_{1}$ derivatives in the
first $\boldsymbol{\mu}$ and $n_{2}$ in the second $\boldsymbol{\mu}$,
the number of distinct permutations is therefore $N!/n_{1}!n_{2}!$
(where $N=n_{1}+n_{2}$ is the total number of derivatives) if $n_{1}\not=n_{2}$
and $N!/2(n_{1}!)^{2}$ otherwise. When inserted in the the expansions
\eqref{eq:exp-1}, \eqref{eq:exp-1-1} the $N!$ factor simplifies away.

To obtain the derivative expansion, we proceed as in Sect. (\ref{matrices}).
The standard Gaussian exponent including the $M$-dependent factor
is\begin{equation}
\frac{1}{2}\mathrm{Tr}(\log M)-\frac{1}{2}[\boldsymbol{m}-\boldsymbol{\mu}(p_{\alpha})]M[\boldsymbol{m}-\boldsymbol{\mu}(p_{\alpha})]\label{eq:orig-1-1}\end{equation}
 Now we expand to second order in the derivatives around the best
fit $\hat{p}_{\alpha}$ not only the $\boldsymbol{\mu}$ term , \begin{equation}
\boldsymbol{\mu}\approx\hat{\boldsymbol{\mu}}+\boldsymbol{\mu}_{,\alpha}\Delta p_{\alpha}+\frac{1}{2}\boldsymbol{\mu}_{,\alpha\beta}\Delta p_{\alpha}\Delta p_{\beta}\end{equation}
 but also $G\equiv\log M$\begin{align}
G & \approx\hat{G}+G_{,\alpha}\Delta p_{\alpha}+\frac{1}{2}G_{,\alpha\beta}\Delta p_{\alpha}\Delta p_{\beta}\\
M & \approx\hat{M}\bigg(1+G_{,\alpha}\Delta p_{\alpha}+\frac{1}{2}G_{,\alpha\beta}\Delta p_{\alpha}\Delta p_{\beta}\bigg)\label{eq:mexp}\end{align}
 where $\hat{M}=M(\hat{p}_{\alpha})$. Then on averaging we obtain

\begin{align}
&\bigg\langle\frac{1}{2}\mathrm{Tr}(\log M)-\frac{1}{2}[\boldsymbol{m}-\boldsymbol{\mu}(p_{\alpha})] M[\boldsymbol{m}-\boldsymbol{\mu}(p_{\alpha})]\bigg\rangle  \approx\nonumber \\
&-\frac{1}{2}F_{\alpha\beta}\Delta p_{\alpha}\Delta p_{\beta}\nonumber \\
&-\frac{1}{2}(\boldsymbol{\mu}_{,\alpha\beta} M\boldsymbol{\mu}_{,\gamma}+\boldsymbol{\mu}_{,\alpha}M_{,\beta}\boldsymbol{\mu}_{,\gamma})\Delta p_{\alpha}\Delta p_{\beta}\Delta p_{\gamma}\nonumber \\
&-\frac{1}{8}(\boldsymbol{\mu}_{,\alpha\beta}M\boldsymbol{\mu}_{,\gamma\delta}+ 4\boldsymbol{\mu}_{,\alpha\beta} M_{,\gamma}\boldsymbol{\mu}_{,\delta}+2\boldsymbol{\mu}_{,\alpha}M_{,\beta\gamma}\boldsymbol{\mu}_{,\delta})\Delta p_{\alpha}\Delta p_{\beta}\Delta p_{\gamma}\Delta p_{\delta}\nonumber \\
&-\frac{1}{4}(\frac{1}{2}\boldsymbol{\mu}_{\alpha\beta} M_{,\gamma}\boldsymbol{\mu}_{\delta\sigma}+\boldsymbol{\mu}_{,\alpha}M_{,\beta\gamma}\boldsymbol{\mu}_{,\delta\sigma})\Delta p_{\alpha}\Delta p_{\beta}\Delta p_{\gamma}\Delta p_{\delta}\Delta p_{\sigma}\nonumber \\
&-\frac{1}{16}\boldsymbol{\mu}_{,\alpha\beta}M_{,\gamma\tau}\boldsymbol{\mu}_{,\delta\sigma}\Delta p_{\alpha}\Delta p_{\beta}\Delta p_{\gamma}\Delta p_{\delta}\Delta p_{\sigma}\Delta p_{\tau}
\end{align}
 The term asymptotically dominant is the last one. It is negative
definite (as required for the normalizability condition) only if $M_{,\alpha\beta}$
is positive definite, i.e. when the covariance matrix is a convex
function of the parameters

\bibliographystyle{mn2e_eprint}
\bibliography{syst-bias}

\begin{thebibliography}{}

\bibitem[\protect\citeauthoryear{{Abramo}}{{Abramo}}{2012}]{Abramo:2012}
{Abramo} L.~R.,  2012, MNRAS, 420, 2042, \eprint{1108.5449},
  \adsurl{http://adsabs.harvard.edu/abs/2012MNRAS.420.2042A}

\bibitem[\protect\citeauthoryear{Akeret, Seehars, Amara, Refregier \&
  Csillaghy}{Akeret et~al.}{2012}]{Akeret:2012ky}
Akeret J.,  Seehars S.,  Amara A.,  Refregier A.,    Csillaghy A.,  2012,
  \eprint{1212.1721}

\bibitem[\protect\citeauthoryear{Albrecht, Bernstein, Cahn, Freedman, Hewitt
  et~al.,}{Albrecht et~al.}{2006}]{Albrecht:2006um}
Albrecht A.,  Bernstein G.,  Cahn R.,  Freedman W.~L.,  Hewitt J.,    et~al.,
  2006, \eprint{astro-ph/0609591}

\bibitem[\protect\citeauthoryear{{Amanullah} et~al.,}{{Amanullah}
  et~al.}{2010}]{AmanullahLidman2010}
{Amanullah} et~al., 2010, Ap.J., 716, 712, \eprint{1004.1711},
  \adsurl{http://adsabs.harvard.edu/abs/2010ApJ...716..712A}

\bibitem[\protect\citeauthoryear{{Amendola} et~al.,}{{Amendola}
  et~al.}{2013}]{Euclid}
{Amendola} L.,  et~al., 2013, Living Reviews in Relativity, 16, 6,
  \eprint{1206.1225},
  \adsurl{http://adsabs.harvard.edu/abs/2013LRR....16....6A}

\bibitem[\protect\citeauthoryear{Amendola, Fogli, Guarnizo, Kunz \&
  Vollmer}{Amendola et~al.}{2014}]{Amendola:2013qna}
Amendola L.,  Fogli S.,  Guarnizo A.,  Kunz M.,    Vollmer A.,  2014,
  Phys.Rev., D89, 063538, \eprint{1311.4765}

\bibitem[\protect\citeauthoryear{Amendola, Marra \& Quartin}{Amendola
  et~al.}{2013}]{Amendola:2012wc}
Amendola L.,  Marra V.,    Quartin M.,  2013, Mon.Not.Roy.Astron.Soc., 430,
  1867, \eprint{1209.1897}

\bibitem[\protect\citeauthoryear{{Amendola} \& {Tsujikawa}}{{Amendola} \&
  {Tsujikawa}}{2010}]{Amendola:2010}
{Amendola} L.,  {Tsujikawa} S.,  2010, {Dark Energy: Theory and Observations},
  \adsurl{http://adsabs.harvard.edu/abs/2010deto.book.....A.
}

\bibitem[\protect\citeauthoryear{Bacon, Goldberg, Rowe \& Taylor}{Bacon
  et~al.}{2006}]{Bacon:2005qr}
Bacon D.~J.,  Goldberg D.,  Rowe B.,    Taylor A.,  2006,
  Mon.Not.Roy.Astron.Soc., 365, 414, \eprint{astro-ph/0504478}

\bibitem[\protect\citeauthoryear{Bassett, Fantaye, Hlozek \& Kotze}{Bassett
  et~al.}{2011}]{Bassett:2009uv}
Bassett B.~A.,  Fantaye Y.,  Hlozek R.,    Kotze J.,  2011, Int.J.Mod.Phys.,
  D20, 2559, \eprint{0906.0993}

\bibitem[\protect\citeauthoryear{{Bueno Belloso}, {Garc{\'{\i}}a-Bellido} \&
  {Sapone}}{{Bueno Belloso} et~al.}{2011}]{2011JCAP...10..010B}
{Bueno Belloso} A.,  {Garc{\'{\i}}a-Bellido} J.,    {Sapone} D.,  2011, \jcap,
  10, 10, \eprint{1105.4825},
  \adsurl{http://adsabs.harvard.edu/abs/2011JCAP...10..010B}

\bibitem[\protect\citeauthoryear{Chevallier \& Polarski}{Chevallier \&
  Polarski}{2001}]{Chevallier:2000qy}
Chevallier M.,  Polarski D.,  2001, Int.J.Mod.Phys., D10, 213,
  \eprint{gr-qc/0009008}

\bibitem[\protect\citeauthoryear{Christensen, Meyer, Knox \& Luey}{Christensen
  et~al.}{2001}]{Christensen:2001gj}
Christensen N.,  Meyer R.,  Knox L.,    Luey B.,  2001, Class.Quant.Grav., 18,
  2677, \eprint{astro-ph/0103134}

\bibitem[\protect\citeauthoryear{{Debono}}{{Debono}}{2013}]{Debono:2013}
{Debono} I.,  2013, MNRAS, 437, 887, \eprint{1308.5636},
  \adsurl{http://adsabs.harvard.edu/abs/2013MNRAS.tmp.2599D}

\bibitem[\protect\citeauthoryear{Dunkley, Bucher, Ferreira, Moodley \&
  Skordis}{Dunkley et~al.}{2005}]{Dunkley:2004sv}
Dunkley J.,  Bucher M.,  Ferreira P.~G.,  Moodley K.,    Skordis C.,  2005,
  Mon.Not.Roy.Astron.Soc., 356, 925, \eprint{astro-ph/0405462}

\bibitem[\protect\citeauthoryear{Feroz \& Hobson}{Feroz \&
  Hobson}{2008}]{Feroz:2007kg}
Feroz F.,  Hobson M.,  2008, Mon.Not.Roy.Astron.Soc., 384, 449,
  \eprint{0704.3704}

\bibitem[\protect\citeauthoryear{Feroz, Hobson \& Bridges}{Feroz
  et~al.}{2009}]{Feroz:2008xx}
Feroz F.,  Hobson M.,    Bridges M.,  2009, Mon.Not.Roy.Astron.Soc., 398, 1601,
  \eprint{0809.3437}

\bibitem[\protect\citeauthoryear{Goldberg \& Bacon}{Goldberg \&
  Bacon}{2005}]{Goldberg:2004hh}
Goldberg D.~M.,  Bacon D.~J.,  2005, Astrophys.J., 619, 741,
  \eprint{astro-ph/0406376}

\bibitem[\protect\citeauthoryear{Heneka, Marra \& Amendola}{Heneka
  et~al.}{2014}]{Heneka:2013hka}
Heneka C.,  Marra V.,    Amendola L.,  2014, Mon.Not.Roy.Astron.Soc., 439,
  1855, \eprint{1310.8435}

\bibitem[\protect\citeauthoryear{Joachimi \& Taylor}{Joachimi \&
  Taylor}{2011}]{Joachimi:2011iq}
Joachimi B.,  Taylor A.,  2011, Mon.Not.Roy.Astron.Soc., 416, 1010,
  \eprint{1103.3370}

\bibitem[\protect\citeauthoryear{{Khedekar} \& {Majumdar}}{{Khedekar} \&
  {Majumdar}}{2013}]{Khedekar:2013}
{Khedekar} S.,  {Majumdar} S.,  2013, JCAP, 2, 30, \eprint{1210.5586},
  \adsurl{http://adsabs.harvard.edu/abs/2013JCAP...02..030K}

\bibitem[\protect\citeauthoryear{{Kosowsky}, {Milosavljevic} \&
  {Jimenez}}{{Kosowsky} et~al.}{2002}]{2002PhRvD..66f3007K}
{Kosowsky} A.,  {Milosavljevic} M.,    {Jimenez} R.,  2002, \prd, 66, 063007,
  \eprint{astro-ph/0206014},
  \adsurl{http://adsabs.harvard.edu/abs/2002PhRvD..66f3007K}

\bibitem[\protect\citeauthoryear{Lewis \& Bridle}{Lewis \&
  Bridle}{2002}]{Lewis:2002ah}
Lewis A.,  Bridle S.,  2002, Phys.Rev., D66, 103511, \eprint{astro-ph/0205436}

\bibitem[\protect\citeauthoryear{Linder}{Linder}{2003}]{Linder_wa}
Linder E.~V.,  2003, Phys. Rev. Lett., 90, 091301, \eprint{astro-ph/0208512}

\bibitem[\protect\citeauthoryear{{Rodriguez}, {Farr}, {Farr} \&
  {Mandel}}{{Rodriguez} et~al.}{2013}]{Rodriguez:2013}
{Rodriguez} C.~L.,  {Farr} B.,  {Farr} W.~M.,    {Mandel} I.,  2013, Phys.
  Rev., D88, 084013, \eprint{1308.1397},
  \adsurl{http://adsabs.harvard.edu/abs/2013PhRvD..88h4013R}

\bibitem[\protect\citeauthoryear{Tegmark, Taylor \& Heavens}{Tegmark
  et~al.}{1997}]{Tegmark:1996bz}
Tegmark M.,  Taylor A.,    Heavens A.,  1997, Astrophys.J., 480, 22,
  \eprint{astro-ph/9603021}

\bibitem[\protect\citeauthoryear{Tegmark \& Zaldarriaga}{Tegmark \&
  Zaldarriaga}{2000}]{Tegmark:2000db}
Tegmark M.,  Zaldarriaga M.,  2000, Astrophys.J., 544, 30,
  \eprint{astro-ph/0002091}

\bibitem[\protect\citeauthoryear{Wang, Percival, Cimatti, Mukherjee, Guzzo
  et~al.,}{Wang et~al.}{2010}]{wang10}
Wang Y.,  Percival W.,  Cimatti A.,  Mukherjee P.,  Guzzo L.,    et~al., 2010,
  Mon.Not.Roy.Astron.Soc., 409, 737, \eprint{1006.3517}

\bibitem[\protect\citeauthoryear{{Wolz}, {Kilbinger}, {Weller} \&
  {Giannantonio}}{{Wolz} et~al.}{2012}]{Wolz:2012}
{Wolz} L.,  {Kilbinger} M.,  {Weller} J.,    {Giannantonio} T.,  2012, JCAP, 9,
  9, \eprint{1205.3984},
  \adsurl{http://adsabs.harvard.edu/abs/2012JCAP...09..009W}

\end{thebibliography}

\label{lastpage} \bsp
\end{document}